\documentstyle[11pt]{article}
\begin{document}
\thispagestyle{empty}
\begin{center}

\vspace{1.8cm} {\bf NON COMMUTATIVE SCALAR FIELDS FROM SYMPLECTIC DEFORMATION}\\
\vspace{1.5cm} {\bf M. Daoud}$^a${\footnote {\it Facult\'e des
Sciences, D\'epartement de Physique,
Agadir, Morocco; email: m$_-$daoud@hotmail.com}} and {\bf A. Hamama}$^b$
\\\vspace{0.5cm}
$^a$ {\it Abdus Salam International Centre for Theoretical Physics,\\
Strada Costiera 11, 34014 Trieste, Italy}\\ \vspace{0.2cm} $^b${\it
High Energy Laboratory, Faculty of Sciences, University Mohamed V,\\
P.O. Box 1014 , Rabat ,
Morocco}\\[1em]

\vspace{3cm} {\bf Abstract}
\end{center}
\baselineskip=18pt
\medskip
This paper is concerned with the quantum theory of noncommutative
scalar fields in two dimensional space time. It is shown that the
noncommutativity originates from the the deformation of symplectic
structures. The quantization is performed and the modes expansions of the fields,
in presence of an electro-magnetic background, are derived. The
Hamiltonian of the theory is given and the degeneracies lifting,
induced by the deformation, is also discussed.

\newpage
\section{Introduction}
In the last decade the noncommutative field theories have been
extensively investigated (for a review see [1-2]) in connection with
the low energy description of string theory with a constant magnetic
background in the presence of $D$-branes [3]. It is important to
note the construction of field theories in noncommutative space-time
encounters many problems as the violation  of Lorentz invariance,
the ultraviolet divergences, and the breaking of unitarity and
causality. However, despite such technical problems, the
noncommutative geometry still considered as one promising candidate
to provide a rigorous framework to investigate and to understand the
non
commutative space time structure at Planck scale [4] in relation with  quantum gravity.\\
It is also important to mention that  motivated by the
noncommutative field theories the quantum mechanical problems in
noncommutative space have received a considerable attention [5-9].
More recently it was shown that the noncommutative geometry via the
Moyal star product is relevant in many branches of physics  and
especially in condensed matter as for instance quantum Hall effect
in different geometries [10-11], bosonization in higher dimensions
[12] and braiding quantum statistics [13-19].\\

In this work, we shall be concerned with the quantum theory of
noncommutative scalar fields where the fields and their conjugate
momenta obey deformed equal time commutations rules. It is important to stress that the quantum theory of
noncommutative fields is defined on a commutative space-time
contrarily  to  the noncommutative field theories which arise from
the non-commutativity of space-time.  The interest for this new kind of quantum
theory of noncommutative fields, introduced in [20-24], is
essentially motivated by Lorentz invariance violation, the problem
of neutrino oscillations and the asymmetry of the dispersion
relation for particles and
antiparticles. \\

In this paper we propose a noncommutative formulation of two
dimensional scalar field theory underlying the noncommutative
algebra with a constant commutator of fields and simultaneously with
a constant commutator of momenta as well. This formulation is done
from a symplectic point of view. Indeed we first suggest a modification of the
symplectic structure associated to phase space associated
 to two massless scalar fields. Secondly, performing a dressing transformation, we
write the deformed two-form in a canonical form. The Hamiltonian is
then converted in a new one involving terms arising from the
deformation. We show that the dynamic described by a
deformed Hamiltonian and canonical two-form is equivalent to the description
which uses undeformed Hamiltonian and deformed symplectic structure. Note
that in a particular case, which will be specified below, the
construction developed here agrees with the results derived in [24]
and most importantly it start at classical level by deforming the
symplectic structure. In quantizing the theory the deformation
induces a modification of the commutation rules as well as in the
spectrum of the system. These results can be viewed as an extension of
the symplectic approach to non-commutative mechanics initiated in [25-27].  \\

The arrangement of this paper is as follows. In the section 2, we
first review some elements concerning symplectic structures in
deriving the evolution of classical fields and the canonical
quantization. This review, included here for completeness, is useful
and necessary to understand and to perform the approach developed in
the next sections. The section 3 concerns the deformation from a
symplectic point of view. We consider two classical scalar fields
and we modify the symplectic two-form to take into account the
presence of an electro-magnetic background. We define the
corresponding Poisson brackets from which we get ones between fields
and their conjugate momenta. In section 4, following the procedure
suggested in [20-24], we quantize the theory,  we propose the
Hamiltonian associated to a quantum theory of two noncommutative
scalar fields and we discuss the degeneracies lifting induced by the
deformation. Solving the Heisenberg equations of motion, we give the
modes expansion of the noncommutative fields. As mentioned above,
the approach essentially originates from the deformation of the
symplectic structures (symplectic two-form and Poisson brackets) in
the phase space of the scalar fields. We end section 4 with  some comments
concerning the  dressing transformation, which play an important
role in the
quantizing the model. Concluding remarks close this paper.\\

\section{Symplectic structures and field quantization}
We shall consider two real massless bosonic fields denoted by
$\Phi^i({\bf x})\equiv \Phi( x , t)$, $i = 1, 2$. The corresponding
action is
\begin{eqnarray}
S = \int_{\Sigma}dtdx {\cal L} =\frac{1}{2l}\sum_{i=1,2}
\int_{\Sigma}dtdx((\partial_t\Phi^i)(\partial_t\Phi^i) -
(\partial_x\Phi^i)(\partial_x\Phi^i)).
\end{eqnarray}
The space-time  region $\Sigma$ will be considered to be of the
form $[0 , l]\times[0, T]$ where $[ 0 , l]$ is line segment is the
spatial region. The equation of motion are given by the
variational principle such that the classical orbit of the fields
which connects the initial and final configurations $\Phi^i( x , 0)$ and $\Phi^i( x , T)$ at time $ t = 0$ and $ t = T$,
minimizes the action. Hence, the minimization condition $\delta S
= 0$ gives the equations of motion (Euler-Lagrange equations)
since the surface contribution of $\delta S$:
\begin{eqnarray}
\delta S_{surface} = \sum_{i=1,2} \int dtdx
\frac{\partial}{\partial x^{\mu}}\bigg(\frac{\partial{\cal
L}}{\partial(\partial_{\mu}\Phi^i)}\delta\Phi^i \bigg)
\end{eqnarray}
vanishes when we fix the initial and final fields configurations,
i.e. $\delta \Phi^i = 0$ at time $t = 0$ and $ t = T$ and also
assume that either $\delta \Phi^i = 0$ or $\frac{\partial {\cal
L}}{\partial(\partial_1\Phi^i)}$ vanishes at $x =0$ and $ x = l$ .
In equation (2) $\mu = 0 ,1$ stands for time and space directions
respectively. The summation over repeated indices is understood.
This is the standard way to get the trajectory of classical fields
by means of Euler-Lagrange equations. There exists another way
based on the symplectic structure of the phase space to describe
the classical evolution of the fields. Indeed, if one consider the
variations of the fields with $\delta \Phi^i$ non-vanishing at
time $t = 0$ and $t = T$, the integration of the surface
contribution term gives \\
\begin{eqnarray}
l\int dtdx \frac{\partial}{\partial
x^{\mu}}\bigg(\frac{\partial{\cal
L}}{\partial(\partial_{\mu}\Phi^i)}\delta\Phi^i \bigg) = {\cal
A}(T) - {\cal A}(0)
\end{eqnarray}
where
\begin{eqnarray}
{\cal A}(t) = \sum_{i=1,2}\int dx
\partial_{t}\Phi^i\delta\Phi^i.
\end{eqnarray}
This is the canonical one form, i.e. a differential one form on
the phase space. In equation (4),  $\delta $ denotes the exterior derivative
on the field space and the time derivatives of the fields must be
treated as variables since ${\cal A}$ is defined at fixed time.
The phase space whose variables (coordinates) are $\Phi^i(x,t)$
and the canonical momentum $\Pi^i(x,t) =
\partial_{t}\Phi^i(x,t)$ is now  equipped  with a
symplectic structure, a canonical two-form, defined by
\begin{eqnarray}
\Omega_0 = \delta{\cal A}(t) = \sum_{i=1,2}\int dx
\delta\Phi^i\wedge\delta\Pi^i
\end{eqnarray}
which can be seen as field strength  corresponding to one-form
${\cal A}$ viewed as $U(1)$ gauge potential. It can also be
written as
\begin{eqnarray}
\Omega_0 = \frac{1}{2}\sum_{I,J}\int dx dx'(\Omega_0)_{IJ}(x,x')
\delta\xi^{I}(x)\wedge\delta\xi^{J}(x')
\end{eqnarray}
where we denote the phase space coordinates by $\xi^I(x) \equiv
\xi^{ii'}(x) = \Phi^i(x)$ (resp. $\Pi^i(x)$) for $i'=1$ (resp.
for $i'=2$) and
\begin{eqnarray}
(\Omega_0)_{IJ}(x,x')\equiv (\Omega_0)_{IJ}\delta(x-x')=
\delta_{ij}\epsilon_{i'j'}\delta(x-x').
\end{eqnarray}
It follows that the Poisson bracket of two functionals ${\cal F}$
and ${\cal G}$ is
\begin{eqnarray}
\{{\cal F} , {\cal G}\} = \sum_{I,J}\int dx
(\Omega_0)^{IJ}\frac{\delta {\cal F}}{\delta\xi^{I}}\frac{\delta
{\cal G}}{\delta\xi^{J}}
\end{eqnarray}
where $\Omega_0^{IJ}$ are the elements of the inverse matrix  of
$\Omega_{IJ}$. In particular the Poisson brackets for the
coordinates on the phase space are the inverse of the symplectic
two-form (treated as matrix):
\begin{eqnarray}
\{\xi^{ik}(x') , \xi^{jl}(x)\} =
\delta^{ij}\epsilon^{kl}\delta(x-x'),
\end{eqnarray}
or alternatively
\begin{eqnarray}
\{ \Phi^i( x , t) , \Phi^j( x' , t)\} = 0{\hskip 0.5cm} \{
\Phi^i( x , t) , \Pi^j( x',t)\} = \delta^{ij}\delta(
x- x'){\hskip 0.5cm}\{ \Pi^i( x , t) , \Pi^j(
x' ,t)\}= 0.
\end{eqnarray}
As we are concerned with two component scalar field theory in a
compact domain, with the boundary condition we can expand the
fields as
\begin{eqnarray}
\Phi^i(  x , t ) = \frac{1}{\sqrt{ l}}\sum_{n \in \mathbf{Z}}
q_n^i\exp(-i n x)
\end{eqnarray}
 and
 \begin{eqnarray}
\Pi^i( x , t ) = \frac{1}{\sqrt{ l}}\sum_{n \in \mathbf{Z}}
p_n^i \exp(-i n x)
\end{eqnarray}
where the normal or Fourier modes $q_n$ and $p_n$ satisfy the
conditions
\begin{eqnarray}
 q_{-n} = q_n^{\star} {\hskip 1.5cm}p_{-n} = p_n^{\star}
\end{eqnarray}
 required by the reality of fields and  their time
 derivatives.\\
Using the equations (5) and (11-12), the symplectic two-form
$\Omega_0$ rewrites
\begin{eqnarray}
\Omega_0 =\sum_{i=1,2} \sum_{n\in {\bf Z} }\delta q^i_n \wedge
\delta p^i_{-n}
\end{eqnarray}
and the Poisson brackets take the simple form
\begin{eqnarray}
\{{\cal F} , {\cal G}\} = \sum_{i=1,2}\sum_{n\in {\bf Z}}
\frac{\delta {\cal F}}{\delta q^{i}_n}\frac{\delta {\cal G}}{\delta
p^{i}_{-n}} - \frac{\delta {\cal F}}{\delta p^{i}_{-n}}\frac{\delta
{\cal G}}{\delta q^{i}_n}.
\end{eqnarray}
In particular, the Poisson brackets corresponding to the canonical
coordinates of the phase space generated by the Fourier modes
$q_n^i$ and $p_n^i$ are given by
\begin{eqnarray}
\{ q_n^i , q_m^j\} = 0 {\hskip 0.5cm} \{ p_n^i , p_m^j\} = 0 {\hskip
0.5cm} \{  q_n^i, p_m^j\} = \delta_{i,j}\delta_{m+n,0}.
\end{eqnarray}
The lagrangian is given by
\begin{eqnarray}
L = \int dx {\cal L} = \frac{1}{2}\sum_{in} (p_n^i p_{-n}^i - n^2
q_n^iq_{-n}^i)
\end{eqnarray}
in terms of the Fourier modes. The canonical Hamiltonian reads
\begin{eqnarray}
H_0 = \sum_{i=1,2}\int dx \Pi^i\partial_t\Phi^i - \int dx {\cal L} =
\frac{1}{2}\sum_{in} (p_n^i p_{-n}^i + n^2 q_n^iq_{-n}^i).
\end{eqnarray}
Since we defined $p^i_n = {\dot q}^i_n$, the canonical momenta
conjugate to $q_n^i$ is defined by
\begin{eqnarray}
\pi_n^i = \frac{\partial L}{\partial\dot{q}_n^i} = \dot{q}_{-n}^i =
p^i_{-n}.
\end{eqnarray}
The Hamiltonian is the generator of time translation. The time
evolution of a function ${\cal F}$ is given by the Hamilton's
equation of motion $\dot{{\cal F}}=\{ {\cal F} , H_0\}$ that gives
\begin{eqnarray}
\frac{dq_n^i}{dt} = \{ q_n^i, H_0\} =  p_{n}^i {\hskip 1cm}
\frac{dp_{n}^i}{dt} = \{ p_{n}^i, H_0\} =-n^2q_{-n}^i
\end{eqnarray}
To pass over the quantum theory in the Heisenberg picture, all
canonical variables become Heisenberg operators satisfying
commutation relations corresponding to Poisson brackets as
\begin{center}
(Poisson Bracket) $\rightarrow$ $-i$ (commutator)
\end{center}
The equations (20) are easily solvable and one recover the
evolution of the fields which can also be derived from the
Euler-Lagrange equations. However, the symplectic prescription
discussed has the merit to be more appropriate to introduce
deformed target space in the spirit of the recent results obtained
in [20-23] and [24]. This issue is the purpose of the next
section.

\section{{Deformed symplectic structure}}

Hereafter, we consider two components abelian field theory in two
dimensional space-time. The generalization to arbitrary number of
components is straightforward. As mentioned above the dynamic of the
system is governed by the Hamiltonian $H_0$ (18). We now assume that
the symplectic structure of the phase space is modified due to the
interaction between the field components $\Phi^1$ and $\Phi^2$ in presence
of an  electromagnetic background. This can be formulated by replacing
the canonical two-form $\Omega_0$ by a closed new one as follows
\begin{eqnarray}
\Omega = \Omega_0 + \frac{1}{2} {\cal E}_{ij} \int dx
\delta\xi^{i2}(x) \wedge\delta\xi^{j2}(x)- \frac{1}{2}{\cal
B}_{ij}\int dx \delta\xi^{i1}(x)\wedge\delta\xi^{j1}(x)
\end{eqnarray}
where the deformation is encoded in the constant antisymmetric
tensors ${\cal E}_{ij}$ and ${\cal B}_{ij}$.\\
Alternatively, $\Omega$ given by (21) can be cast in a compact form
\begin{eqnarray}
\Omega = \frac{1}{2}  \int dxdx' \Omega_{IJ}(x,x') \delta\xi^{I}(x)
\wedge\delta\xi^{J}(x')
\end{eqnarray}
providing the non trivial deformed metric
\begin{eqnarray}
\Omega_{IJ}(x,x') = \Omega_{IJ}\delta (x-x')
\end{eqnarray}
with
\begin{eqnarray}
\Omega_{IJ} = \delta_{ij}\epsilon_{i'j'} + {\cal
E}_{ij}\delta_{i'2}\delta_{j'2} - {\cal
B}_{ij}\delta_{i'1}\delta_{j'1}
\end{eqnarray}
which is non degenerate (det$\Omega \neq 0$) when the antisymmetric
tensors ${\cal E}_{ij}$ and ${\cal B}_{ij}$ satisfy the condition
$\det( 1_{2\times 2} - {\cal E}{\cal B}) \neq 0$. This is easily
seen in writing $\Omega$ in a matrix form. We assume in this work
that such a condition is satisfied. To find the classical equations
of motion and to establish the connection between the classical and
quantum theory, it is necessary to define the Poisson brackets
associated with the new phase space geometry in a consistent way.
Indeed, recalling that the Poisson brackets for the coordinates on
the phase space are the inverse of the symplectic form as matrix, we
have
\begin{eqnarray}
\{{\cal F} , {\cal G}\} = \sum_{I,J}\int dx \Omega^{IJ}\frac{\delta
{\cal F}}{\delta\xi^{I}}\frac{\delta {\cal G}}{\delta\xi^{J}}
\end{eqnarray}
where $\Omega^{IJ}$ is the inverse matrix of $\Omega_{IJ}$ (24) and
${\cal F}$ and ${\cal G}$ are two functionals on the phase space. \\
As we are interested by scalar field theory, we use the
equations (11-12) and (22) to write the symplectic form  as
\begin{eqnarray}
\Omega = \delta q^{i}_{n}\wedge\delta p^{i}_{-n}+\frac{1}{2}{\cal
E}_{ij}\delta p^{i}_{n}\wedge\delta p^{j}_{-n}- \frac{1}{2}{\cal
B}_{ij}\delta q^{i}_{n}\wedge\delta q^{j}_{-n}
\end{eqnarray}
and the
Poisson brackets take the simple form
\begin{eqnarray}
\{{\cal F}, {\cal G}\} = \sum_{ikn} (\omega^{-1}_1)_{ik}\frac{\delta
{\cal F}}{\delta q^i_n}\bigg[ \frac{\delta {\cal G}}{\delta
p^k_{-n}} - \sum_{j} {\cal E}_{kj}\frac{\delta {\cal G}}{\delta
q^j_{-n}} \bigg] -  (\omega^{-1}_2)_{ik}  \frac{\delta {\cal
F}}{\delta p^i_{-n}}\bigg[ \frac{\delta {\cal G}}{\delta q^k_n}-
\sum_j{\cal B}_{kj}\frac{\delta {\cal G}}{\delta p^j_n}\bigg]
\end{eqnarray}
where the functionals ${\cal F}$ and ${\cal G}$ are now expressed in
terms of $q^i_n$ and $p^i_n$ generating the phase space. The
matrix elements of $\omega_1$ and $\omega_2$, occurring in (27), are defined by
\begin{eqnarray}
(\omega_1)_{ij} = \delta_{ij} - {\cal E}_{ik}{\cal B}_{kj},
\end{eqnarray}
\begin{eqnarray}
(\omega_2)_{ij} = \delta_{ij} - {\cal B}_{ik}{\cal E}_{kj}.
\end{eqnarray}
The last equations can be read in matrices form as $\omega_1 = 1 -
{\cal E}{\cal B}$ and $\omega_2 = 1 - {\cal B}{\cal E}$,
respectively. It follows that the modified canonical Poisson
brackets are
\begin{eqnarray}
\{ q^i_n , q^j_m\} = - \sum_k(\omega^{-1}_1)_{ik}{\cal
E}_{kj}\delta_{m+n,0},
\end{eqnarray}
\begin{eqnarray}
\{ p^i_n , p^j_m\} =  \sum_k(\omega^{-1}_2)_{ik}{\cal
B}_{kj}\delta_{m+n,0},
\end{eqnarray}
\begin{eqnarray}
\{ q^i_n , p^j_{-m}\} =  (\omega^{-1}_1)_{ij}\delta_{n,m} =
(\omega^{-1}_2)_{ji}\delta_{n,m}.
\end{eqnarray}
According to the modification of the symplectic structure of the phase
space, we introduce the vector fields $X_{\cal F}$ associated to a
given functional  ${\cal F}(q^i_n , p^i_n)$
\begin{eqnarray}
X_{\cal F} = \sum_{in} X^i_n \frac{\delta }{\delta q^i_n} +
Y^i_n\frac{\delta }{\delta p^i_{-n}}
\end{eqnarray}
such that the interior contraction of $\Omega$ with  $X_{\cal F}$
gives
\begin{eqnarray}
{\it i}(X_{\cal F}) \Omega = \delta {\cal F}.
\end{eqnarray}
A straightforward calculation leads to
\begin{eqnarray}
 X^i_n= \sum_j(\omega^{-1}_1)_{ij}\bigg(\frac{\delta {\cal
F}}{\delta p^j_{-n}} - \sum_k{\cal E}_{jk} \frac{\delta {\cal F}
}{\partial q^k_{-n}}\bigg)
\end{eqnarray}
\begin{eqnarray}
 Y^i_n=- \sum_j(\omega^{-1}_2)_{ij}\bigg(\frac{\delta {\cal
F}}{\delta q^j_{n}} - \sum_k{\cal B}_{jk} \frac{\delta {\cal F}
}{\partial p^k_{n}}\bigg)
\end{eqnarray}
 and one can check that
\begin{eqnarray}
{\it i}(X_{\cal F}){\it i}(X_{\cal G}) \Omega = \{{\cal F}, {\cal
G}\}.
\end{eqnarray}
Thus in the deformed case the fields and their conjugate momentum
satisfy the following Poisson algebra
\begin{eqnarray}
\{ \Phi^i( x , t) , \Phi^j( x',t)\} = -
\sum_k(\omega^{-1}_1)_{ik}{\cal E}_{kj}\delta(x - x'),
\end{eqnarray}
\begin{eqnarray}
 \{ \Phi^i( x ,
t) , \Pi^j( x' ,t)\} =  (\omega^{-1}_2)_{ij}\delta( x -
x'),
\end{eqnarray}
\begin{eqnarray}
\{ \Pi^i( x , t) , \Pi^j(  x' ,t)\}=
\sum_k(\omega^{-1}_2)_{ik}{\cal B}_{kj}\delta( x - x').
\end{eqnarray}
Clearly, in the limiting case ${\cal E} = 0$ and ${\cal B} = 0$, we
recover the canonical Poisson brackets (10).\\
To simply our purpose, let us now set
\begin{eqnarray}
{\cal E}_{ij} = \theta {\epsilon}_{ij} {\hskip 1.5cm}{\cal B}_{ij} =
\bar{\theta}{\epsilon}_{ij}
\end{eqnarray}
where $\epsilon_{ij}$ is the usual antisymmetric tensor
($\epsilon_{12} = - \epsilon_{21} = 1$). With this choice, the
Poisson brackets (30-32) read simply
\begin{eqnarray}
\{ q^i_n , q^j_m\} = -
\frac{\theta}{1+\theta\bar{\theta}}\epsilon_{ij}\delta_{m+n,0}
\end{eqnarray}
\begin{eqnarray}
\{ p^i_n , p^j_m\} =
\frac{\bar{\theta}}{1+\theta\bar{\theta }}\epsilon_{ij}\delta_{m+n,0}
\end{eqnarray}
\begin{eqnarray}
\{ q^i_n , p^j_{-m}\} =
\frac{1}{1+\theta\bar{\theta }}\delta_{ij}\delta_{n,m}
\end{eqnarray}
reflecting a deviation from the canonical brackets. At this stage,
it is remarkable that the symplectic form (26) and  the
corresponding Poisson brackets (27) can be converted in the canonical
forms by mean of the so-called dressing transformation  which
furnishes  a simply way to quantize the theory. This is the main
task of the next section.
\section{ Dressing transformation and Quantization}
To begin note that under the following transformation
\begin{eqnarray}
Q^i_n = a q^i_n + \frac{1}{2} b\theta\sum_k\epsilon_{ki} p^k_{n}
\end{eqnarray}
\begin{eqnarray}
P^i_n = c p^i_n + \frac{1}{2} d\bar{\theta}\sum_k\epsilon_{ki}
q^k_{n},
\end{eqnarray}
the Poisson brackets (42-44) give the canonical ones
\begin{eqnarray}
\{ Q^i_n , Q^j_m\} =  0  {\hskip 1.5cm } \{ P^i_n , P^j_m\} = 0
{\hskip 1.5cm } \{ Q^i_n , P^j_{-m}\} = \delta_{ij}\delta_{n,m}
\end{eqnarray}
when the scalars $a$, $b$, $c$ and $d$ satisfy the following
constraints
$$
4a^2 - 4ab-\theta\bar{\theta}b^2 = 0
$$
$$
4c^2 - 4cd-\theta\bar{\theta}d^2 = 0
$$
$$
4ac + 2\theta\bar{\theta}(ad+bc)-\theta\bar{\theta}bd = 4(1 +
\theta\bar{\theta}).
$$
The transformation (45-46) generalizes the so-called dressing transformation introduced in [24].
As  simple solution of the above equations, we consider
\begin{eqnarray}
a = c = \frac{1}{b} = \frac{1}{d}=
\frac{1}{\sqrt{2}}\sqrt{1+\sqrt{1+\theta\bar{\theta}}}.
\end{eqnarray}
With these new dynamical variables, $\Omega$ takes a canonical form
\begin{eqnarray}
\Omega = \sum_{in} \delta Q^i_n \wedge \delta P^i_{-n}.
\end{eqnarray}
Inverting the above dressing transformation (45-46)
\begin{eqnarray}
q^i_n = \frac{a}{\sqrt{1+\theta{\bar\theta}}}\bigg[ Q^i_n +
\frac{\theta}{2a^2} \sum_k\epsilon_{ik} P^k_{n}\bigg]
\end{eqnarray}
\begin{eqnarray}
p^i_n = \frac{a}{\sqrt{1+\theta{\bar\theta}}}\bigg[P^i_n +
\frac{\bar{\theta}}{2a^2}\sum_k\epsilon_{ik} Q^k_{n}\bigg],
\end{eqnarray}
the Hamiltonian (18), denoted in what follows by $H$ to avoid any
confusion, becomes
\begin{eqnarray}
H = \frac{a^2}{2(1+\theta{\bar\theta)}}\bigg[ \sum_{in} (1 +
\frac{\theta^2}{4a^4}n^2) P^i_nP^i_{-n} +
(\frac{\bar{\theta}^2}{4a^4} + n^2)Q^i_nQ^i_{-n} +
(\frac{\theta}{a^2}n^2 + \frac{\bar{\theta}}{a^2})
\sum_j\epsilon_{ij}P^i_{-n}Q^j_n\bigg].
\end{eqnarray}
It is important to note that thanks to the dressing transformation
(45-46), the dynamic described by the Hamiltonian $H_0$ on the
deformed symplectic space is converted to one described by
underformed (canonical) symplectic structure (49) and the Hamiltonian
(52) expressed in terms of the new dynamical modes of the theory.
Evidently the ($\theta ,\bar{\theta}$)-dependent terms in $H$ arise
from the deformation of the symplectic structure. This suggests that
the fields $\Phi^1$ and $\Phi^2$ interacts with a given potential $
V_{int} $
 to have $H = H_0 + V_{int}$ where $H_0$ is the free Hamiltonian
given by (18) modulo the substitution $(q,p) \rightarrow (Q,P) $. In
this respect the deformation of the symplectic structure  can be
seen as a perturbation reflecting the action of  some external
potential on the system. This feature is very similar to the Landau problem in quantum mechanics.\\

We now come to the quantization of the model. Since, as mentioned
early, the two descriptions are equivalent, we shall quantize the
system using the canonical prescription. Thus, we replace the phase
space variables by operators satisfying commutation rules where the
commutators are given by $i$ times the associated Poisson brackets.
Consequently, we have
\begin{eqnarray}
[Q^i_n , Q^j_m] = 0 {\hskip 1cm}[P^i_n , P^j_m]= 0 {\hskip
1cm}[Q^i_n , P^j_{-m}]= i\delta_{m+n,0}\delta_{i,j}.
\end{eqnarray}
This gives
\begin{eqnarray}
[q^i_n , q^j_m]=
-i\frac{\theta}{1+\theta\bar{\theta}}\epsilon_{ij}\delta_{m+n,0}
\end{eqnarray}
\begin{eqnarray}
 [p^i_n , p^j_m] = i\frac{\bar{\theta}}{1+\theta\bar{\theta}}\epsilon_{ij}\delta_{m+n,0}
\end{eqnarray}
\begin{eqnarray}
[q^i_n , p^j_{-m}]=
i\frac{1}{1+\theta\bar{\theta}}\delta_{i,j}\delta_{n,m},
\end{eqnarray}
from which we obtain the equal time commutation relations for the
fields $\Phi^1$, $\Phi^2$ and their conjugate momenta:
\begin{eqnarray}
[\Phi^i( x , t) , \Phi^j( x',t) ] = -i
\frac{\theta}{1+\theta\bar{\theta}}\epsilon_{ij} \delta(
x- x')
\end{eqnarray}
\begin{eqnarray}
[\Pi^i( x ,t) , \Pi^j( x' , t) ] = i
\frac{\bar{\theta}}{1+\theta\bar{\theta}}\epsilon_{ij}
\delta( x -  x')
\end{eqnarray}
\begin{eqnarray}
[ \Phi^i( x ,t ), \Pi^j( x' , t)] = - i
\frac{1}{1+\theta\bar{\theta}}\delta_{ij}\delta( x -
x').
\end{eqnarray}
At this stage, we introduce the operators
\begin{eqnarray}
a^i_n = \sqrt{\frac{\Delta_n}{2}} (Q^i_n + i
\frac{P^i_{n}}{\Delta_n}) {\hskip 1cm} a^{i+}_n =
\sqrt{\frac{\Delta_n}{2}} (Q^i_{-n} - i \frac{P^i_{-n}}{\Delta_n})
\end{eqnarray}
where
\begin{eqnarray}
\Delta_n = \sqrt{\frac{\bar{\theta}^2 + 4a^4n^2}{4a^4 +
\theta^2n^2}}.
\end{eqnarray}
They satisfy the usual Heisenberg commutation relations $[a^i_n ,
a^j_m] = \delta_{i,j}\delta_{m,n}$. The Hamiltonian $H$ can be
written as the sum of two contributions:
\begin{eqnarray}
H-H_L = \frac{1}{4}\frac{1}{1+\theta\bar{\theta}}\sum_{in}
\sqrt{(\bar{\theta}^2 + 4a^4
n^2)(4a^4+\theta^2n^2)}(1+2a^{i+}_na^i_n)
\end{eqnarray}
and
\begin{eqnarray}
H_L = \frac{i}{2}\frac{1}{1+\theta\bar{\theta}}\sum_{ijn}
(\bar{\theta} + \theta n^2) \epsilon_{ij}a^{i+}_na^j_n.
\end{eqnarray}
The Hamiltonian $H$ can be diagonalized by considering the operators
\begin{eqnarray}
A^1_n = \frac{1}{\sqrt{2}}(a^1_n -i a^2_n) {\hskip 1cm}A^2_n =
\frac{1}{\sqrt{2}}(a^1_n +i a^2_n).
\end{eqnarray}
Indeed, substituting (64) in (62-63), we obtain
\begin{eqnarray}
H = \sum_n  [\omega_n + (\omega_n - \bar{\omega}_n)A^{1+}_nA^1_n  +
(\omega_n + \bar{\omega}_n)A^{2+}_nA^2_n ]
\end{eqnarray}
where
\begin{eqnarray}
\omega_n = \frac{1}{2}\frac{1}{1+\theta\bar{\theta}}
\sqrt{(\bar{\theta}^2 + 4a^4 n^2)(4a^4+\theta^2n^2)}
\end{eqnarray}
and
\begin{eqnarray}
\bar{\omega}_n = \frac{1}{2(1+\theta\bar{\theta})}
(\bar{\theta} + \theta n^2)
\end{eqnarray}
From the last three equations, it is easily seen that the
deformation induces a lifting of the degeneracies of the spectrum.
The Hamiltonian is a superposition of two independents one
dimensional oscillator unlike to undeformed case where the
Hamiltonian is a sum of two-dimensional oscillators. The dynamics of
this system is described by the Heisenberg equations:
\begin{eqnarray}
\frac{dA^1_n}{dt} = -i[ A^1_n , H ] = -i \omega_n^- A_n^1 {\hskip
1cm}\frac{dA^2_n}{dt} = -i[ A^2_n , H ] = -i \omega_n^+ A_n^2
\end{eqnarray}
where $\omega^{\pm}_n = \omega_n \pm {\bar{\omega}}_n $. Thus, we
have:
\begin{eqnarray}
A^1_n(t) = \hat{A}_n^1\exp(-i \omega_n^-t)  {\hskip 1cm}A^1_n(t) =
\hat{A}_n^2\exp(-i \omega_n^+t)
\end{eqnarray}
where the operators $\hat{A}^1_n$ and $\hat{A}^2_n$ are
time-independents. Consequently, using the equations (50), (60), (64) and
(69) we obtain the normal modes of the model as
\begin{eqnarray}
q^1_n(t) = \frac{1}{2}\bigg[\Lambda_n^+ \bigg(
\hat{A}^1_{n}\exp(-i\omega_n^-t) +
\hat{A}^{1+}_{-n}\exp(+i\omega^-_nt)\bigg) + \Lambda_n^- \bigg(
\hat{A}^2_{n}\exp(-i\omega_n^+t) +
\hat{A}^{2+}_{-n}\exp(+i\omega_n^+t)\bigg)\bigg]
\end{eqnarray}
 and
\begin{eqnarray}
q^2_n(t) = \frac{i}{2}\bigg[\Lambda_n^+ \bigg(
\hat{A}^1_{n}\exp(-i\omega_n^-t) -
\hat{A}^{1+}_{-n}\exp(+i\omega^-_nt)\bigg) + \Lambda_n^- \bigg(
\hat{A}^{2+}_{-n}\exp(+i\omega_n^+t)-\hat{A}^2_{n}\exp(-i\omega_n^+t)\bigg)\bigg]
\end{eqnarray}
where the $(\theta , \bar{\theta})$-dependent constants
$\Lambda_n^{\pm}$ are defined by
\begin{eqnarray}
\Lambda_n^{\pm} = \bigg[\frac{1}{\sqrt{\Delta_n}} \pm
\frac{\theta}{2a^2}\sqrt{\Delta_n}\bigg].
\end{eqnarray}
It is clear that in the limiting case $\theta = 0$ and $\bar{\theta}
= 0$, one recovers the usual results arising from the equations
(20). Note also that for $\bar{\theta} = 0$, our results agree with ones obtained in [24].
 We end up this section by some remarks related  the dressing transformation (45-46).
 In this sense, we will  show that the deformed Poisson algebra Eqs.(42-44) is
un-equivalent to un-deformed one (47). The un-equivalency occurs if
the two algebras can not be transformed to each other by a unitary
transformation. This means that the dressing transformation (45-46)
should be not unitary or more precisely not orthogonal since its
elements are reals. In fact , it is easy to check that the
orthogonality  requires the following conditions
\begin{eqnarray}
a^2 + \frac{1}{4}\theta^2 b^2 = 1 {\hskip 1cm} c^2 +
\frac{1}{4}{\bar{\theta}}^2 d^2 = 1  {\hskip 1cm} bc \theta = ad
\bar{\theta}.
\end{eqnarray}
Setting
\begin{eqnarray}
 a = \cos\varphi_1{\hskip 0.5cm} c =  \cos\varphi_2    {\hskip 0.5cm} b\theta = 2  \sin\varphi_1 {\hskip 0.5cm}
 d\bar{\theta} = 2  \sin\varphi_2
\end{eqnarray}
one can see that last equality in (73) is satisfied if
\begin{eqnarray}
  \varphi_1 = \varphi_2 + n\pi {\hskip 1cm} n\in \textbf{N}
\end{eqnarray}
which implies
\begin{eqnarray}
a = \pm c {\hskip 1cm} b\theta = \pm d \bar{\theta}
\end{eqnarray}
Consistency with (48) gives $\theta = \bar{\theta}$. This shows that
the dressing transformation is orthogonal when the strengths of
magnetic and electric backgrounds are equal. Finally, we stress that the transformation
(45-46) is similar to Darboux coordinates transformation (see for instance [26-27]).\\

\section{Concluding remarks }
We have clarified a procedure generating the noncommutative scalar
field theories. The key point of this procedure is the deformation
of the symplectic structure of the phase space of classical fields.
Having constructed the Poisson brackets, we quantized the model
under consideration following the standard canonical scheme thanks
to the so called dressing transformation(45-46).\\
An interesting feature of this approach lies on the deformation of
the symplectic structure which is introduced before the quantization
process.  The present results can be extended in various directions.
In particular, we believe that this approach can be adapted to the
theory of noncommutative chiral fields in two dimension in relation
with fractional quantum Hall effect. Indeed, it is well established
that for an incompressible quantum Hall droplet, the edges
excitations are described by a chiral scalar field. In this respect,
the quantum theory of noncommutative chiral fields
can provides us with an unified scheme to classify different Hall hierarchies and
 can bring new fractional filling factors with interesting physical consequences. We hope to report on this issue in a
forthcoming work.

{\vskip 1cm}
 \noindent {\bf Acknowledgments}\\
MD would like to thank the hospitality and kindness of Condensed
 Matter and Statistical Physics section of Abdus Salam International
 Centre for Theoretical Physics (AS-ICTP) where this work was
 done. The authors are indebted to the referee for his
 constructive comment.

\end{document}